\documentclass[a4paper,11pt,fleqn]{article}
\pdfoutput=1 

\usepackage{jheppub} 
\usepackage{xcolor}
\usepackage{natbib}
\setcitestyle{square, comma, numbers,sort&compress, super}
\usepackage{physics}
\usepackage{fancyhdr}
\usepackage{bm,amsmath,bbm,amsfonts,nicefrac,latexsym,amsmath,amsfonts,amsbsy,amscd,amsxtra,amsgen,amsopn,bbm,amsthm,amssymb}
\usepackage[nottoc,numbib]{tocbibind}
\usepackage[export]{adjustbox}
\usepackage{color}
\usepackage{accents}
\usepackage[normalem]{ulem}
\usepackage{graphicx}
\usepackage[colorinlistoftodos]{todonotes}
\usepackage{etoolbox}
\apptocmd{\thebibliography}{\raggedright}{}{}

\usepackage[colorlinks=true,linkcolor=blue]{hyperref}

\title{Bootstrapping PT symmetric Hamiltonians}

\author{Sakil Khan,}
\author{Yuv Agarwal,}
\author{ Devjyoti Tripathy,}
\author{ Sachin Jain}

\affiliation{Indian Institute of Science Education and Research, Homi Bhabha Rd, Pashan, Pune 411 008, India}

\emailAdd{sakil.khan@students.iiserpune.ac.in }
\emailAdd{yuv.agarwal@students.iiserpune.ac.in}
\emailAdd{devjyoti.tripathy@students.iiserpune.ac.in}
\emailAdd{sachin.jain@iiserpune.ac.in}

\abstract{ Bootstrapping in Quantum Mechanics uses positivity condition to derive the Eigen spectum. For non-hermitian systems usual positivity condition does not work. In this paper we define positivity condition for special class of non-hermitian hamiltonian, the PT symmetric Hamiltonian. We illustrate this modified positivity condition with several examples and obtain eigen spectrum.}

\begin{document} 
	\maketitle
	\flushbottom
	
	\pagebreak
	\section{Introduction}
	One of the main tool that one uses to solve quantum mechanical systems is perturbation theory. However perturbation theory can not be used to strongly coupled systems. For this cases, one uses numerical techniques to solve Schrodinger equations. Alternatively one can use Bootstrap techniques, proposed recently in \cite{hartnol} to solve for Eigen values. Important aspect of this technique is that one does not need to solve for Schrodinger equation, rather one imposes some fundamental constraints such as positivity of the norm consistent with symmetries of the system. In principle since this procedure requires no further information than symmetries and positivity of the norm, traditionally it has been used to search consistent theories as well as its spectrum \cite{El-Showk:2012cjh}. However, in the context of quantum mechanics, since there is no restriction on kind of potential that one can write down, the main focus has been to obtain the spectrum of a given quantum mechanical system \cite{Aikawa:2021eai,Aikawa:2021qbl,Bai:2022yfv,Berenstein:2021dyf,Bhattacharya:2021btd,Du:2021hfw,Kazakov:2021lel,Lawrence:2021msm,Nakayama:2022ahr,Tchoumakov:2021mnh,Li:2022vzn}.
	
	Our aim in this paper is to generalise the bootstrapping technique to non-hermitian systems where the positivity of norm can not be demanded. We focus on PT symmetric potentials where the spectrum can be shown to be real \cite{benderprl,ptbender}. PT symmetric Hamiltonian naturally appears in various physically relevant situations such as in studies of
	the Lee-Yang edge singularity \cite{PhysRevLett.40.1610,CARDY1989275}. In this paper we show that with some modified definition of norm for the PT symmetric system, one can impose positivity constraint and obtain the spectrum. This article is organised as follows. In section \ref{bt1} we review the bootstrapping technique for hermitian hamiltonians and solve for exactly solvable model Poschel-Teller potential and discuss number of interesting observations. In section \ref{bt2} we set up bootstrapping technique for PT symmetric Hamiltonian and discuss several examples in section \ref{expt1} to illustrate the point. In section \ref{disc1} we present discussions and future directions. Section \ref{detail} discusses some of the details that is required in the main text.
	
	Note added: Just before posting this paper,  \cite{Li:2022vzn} appeared in ArXiv which also deals with non-hermitian systems. However the approach is different than one employed in this paper.

	\section{Quantum Bootstrapping}\label{bt1}
	
	In a recent paper\cite{hartnol} presented an elegant way of solving the energy Eigen spectrum of a hermitian Hamiltonian. They showed that
	using the following constraints we can obtain the Eigen spectrum numerically of any
	hermitian Hamiltonian,
	\begin{align}\label{eq}
		& \braket{ R_{n}}{[H,O] |R_{n}} =0\nonumber\\
		&\braket{ R_{n}}{HO |R_{n}} =E_{n}\braket{ R_{n}}{O |R_{n}}\nonumber\\
		&    \braket{ R_{n}}{O^\dagger O|R_{n}}\geq 0
	\end{align}
	where O is an arbitrary operator, $\ket{R_{n}}$ is the n’th energy eigenstate and $\bra{R_{n}}$ is the complex conjugate of $\ket{R_{n}}$. The last constraint of the above equation is the positivity condition of state
	i.e. the norm of any state must be positive. There are many examples that has been solved in the literature \cite{hartnol,Aikawa:2021eai,Aikawa:2021qbl,Bai:2022yfv,Berenstein:2021dyf,Bhattacharya:2021btd,Du:2021hfw,Kazakov:2021lel,Lawrence:2021msm,Nakayama:2022ahr,Tchoumakov:2021mnh,Li:2022vzn} using this technique. As an illustration we use this technique to solve exactly solvable model, the Poschel-Teller potential.
	
	\subsection*{Poschl-Teller potential}
	This is an exactly solvble potential, $V(x)=-\frac{\lambda(\lambda+1)}{2}\sech^{2}{x}$. Putting this into the Schrodinger equation and substituting $u=\tanh{x}$, we get the following differential equation 
	\begin{equation}
		\left((1-u^{2})\psi'(u)\right)'+\lambda(\lambda+1)\psi(u)+\frac{2E}{1-u^{2}}\psi(u)=0  
	\end{equation}
	The solution to this differential equation is given by the Legendre functions \begin{equation}
		\psi(u)=P^{\mu}_{\lambda}(\tanh{x}).
	\end{equation} The energy eigenvalues are given by $E=-\frac{\mu^{2}}{2}$ where $\lambda=1,2,3,....., \text{and }  \mu=1,2,3,....,\lambda-1,\lambda$.
	We will use bootstrap technique to find out the spectrum for this potential. To do this, we choose $\mathcal{O}=\sech^{n}{x}\tanh{x}$ to compute 
	\begin{equation}\label{stp1}
		\langle [\mathcal{H},\mathcal{O}] \rangle,~~~~ \langle\mathcal{H}\mathcal{O}\rangle=E\langle \mathcal{O}\rangle.   \end{equation} Using these two equations in \eqref{stp1}, we are able to write down a recursion relation as follows
	\begin{align}\label{rec1}
		&(2tE+\frac{t^3}{2})\langle\sech^{t}{x}\rangle+\big(\frac{t^3}{2}+3t^2+4t+2+\frac{3t+2}{2}-2\lambda(\lambda+1)(t+2)\big)\langle\sech^{t+2}{x}\rangle\nonumber\\
		&+\big(-2(t+1)E+2\lambda(\lambda+1)(t+1)-t^3-3t^2-4t-2\big)\langle\sech^{t+4}{x}\rangle=0
	\end{align}
	
\begin{figure}[h!]
		\centering
		\includegraphics[scale=0.3]{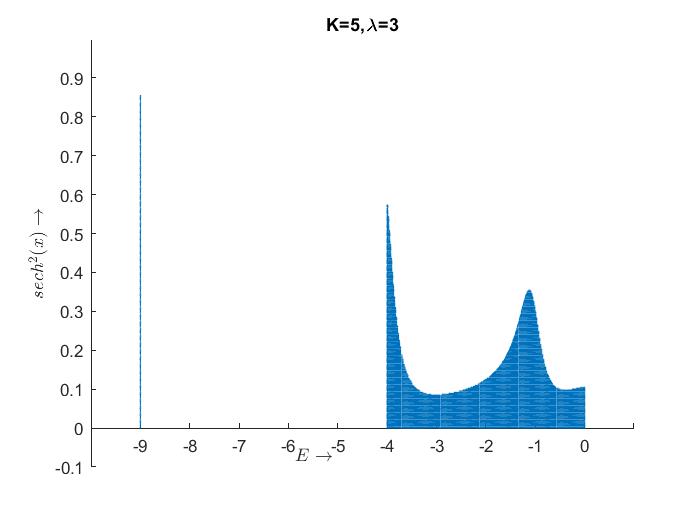} 
		\includegraphics[scale=0.3]{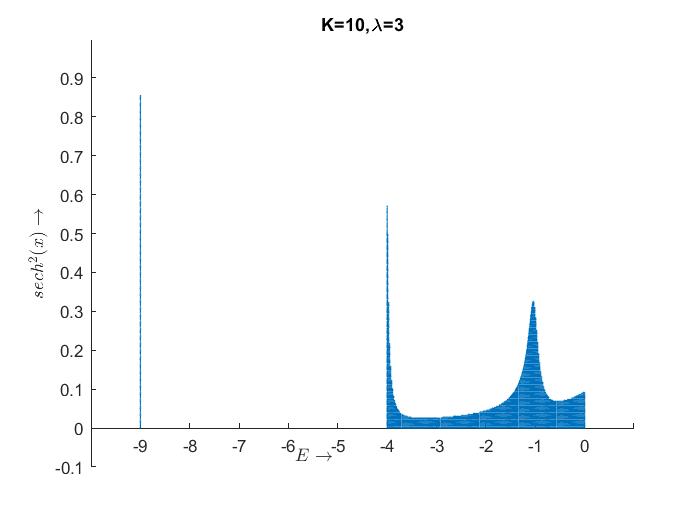}
		\caption{Here we plot the results of bootstrap for $H=p^{2}-\frac{\lambda(\lambda+1)}{2}\sech^{2}{x}$} 
		\label{fig:plot1}
	\end{figure}

	This recursion relation is used to obtain $\langle \sech^{2t}{x}\rangle$ in terms of $\langle \sech^{2}{x}\rangle$ and energy E. This allows us to form a matrix $\mathcal{M}$ given by $\mathcal{M}_{ij}=\sech^{2(i+j)}{x}$. We impose the condition that $\mathcal{M}$ be positive semi-definite to get the energy spectrum. The Eigen spectrum is plotted in Fig\ref{fig:plot1}.

	\newpage
	For the ease of understanding, we have redefined the energy eigenvalues to be E=$-\mu^{2}$. As expected, we find that for $\lambda=3$, there are 3 peaks at E=-1,-4 and -9 corresponding to $\mu=1,2,3$ respectively. K here refers to the size of the matrix M on which positivity constraint is applied. When we compute expectation values of higher powers of $\sech^{2}{x}$ using the recursion relation, the size of the matrix M becomes larger, hence K also increases. As expected, with increase in K (from K=5 to K=10), the peaks become sharper. We also notice that even if the energy values become very localised and occur at their expected positions, there is a very wide spread in the allowed values of $\langle\sech^{2}{x}\rangle$ and this feature does not vanish even when we increase K. We wish to address this issue in future.

	One would expect that it is possible to, at least write down a closed form of recursion relation for hermitian potentials that are exactly solvable like in the case of \eqref{rec1}. However, it turns out that this is not the case. For Poschl Teller potential, we were able to write down the recursion relation in terms of expectation values of even powers of $\sech{x}$. If we consider the potential of the form $V=\sech^{2}{x}+\sech{x}\tanh{x}$, which is also exactly solvable,  we find that its not possible to write down a closed form of recursion relation \footnote{In this case we get odd powers of hyperbolic trigonometric functions which cannot be written back in terms of $\sech^{2}{x}$. In general this will lead to much larger class of search space and hence complicating the problem. To tackle this kind of problems, we may need to impose more general positivity condition. We discuss one such generalisation for two dimensional systems where also obtaining a closed form recursion relation is problematic.}.

	However, this method is limited to the
	hermitian Hamiltonian case only. In \cite{ptbender}, it was shown that   PT-symmetric non-hermitian Hamiltonian can have real eigenvalues, so an important question is then whether it is possible to generalize the ”Bootstrapping” method for PT-symmetric case or not. We discuss this issue in the next section.

	\section{Bootstrapping for PT symmetric non-hermitian Hamiltonian}\label{bt2}
	PT symmetric hamiltonians are important class of Hamiltonians which appears in variety of physical situations. They are interesting because eventhough the Hamiltonian is not hermitian, its eigen values are real. One can solve them either analytically or numerically using Schrodinger equations. In this section we develop bootstrapping technique for solving PT symmetric systems. The first challenge in the PT-symmetric Hamiltonian case is to find the positivity condition. In this section, we are going to obtain a suitable positivity condition.
	
	\subsection{Bootstrapping condition for PT symmetric systems}
	For hermitian system the the left and right eigenvectors are the same and one can define orthonormality condition using them.
	However, in non-hermitian case it is little non-trivial to define orthonormality condition because in this case we have to define a Hamiltonian dependent inner product. Let $\ket{R_{n}}$ is the n'th eigen state i.e.
	\begin{equation}
		H\ket{R_{n}}=E_{n}\ket{R_{n}}
	\end{equation}
	If we define $\bra{R_{n}}$ as the complex conjugate of $\ket{R_{n}}$, then unlike in the case of hemitian systems,  for PT symmetric cases $\braket{R_{m}}{R_{n}} \neq \delta_{mn}$ that 
	is they are not orthonormal.  In \cite{V} introduced a new kind of norm "V norm" which is both positive and orthonormal,
	\begin{equation}\label{eqnho}
		\braket{R_{m}}{V|R_{n}}=\delta_{n,m}
	\end{equation}
	where V operator satisfies the following properties,
	\begin{align}
		&1. VH V^{-1}=H^\dagger\nonumber\\
		&2. V^\dagger=V\nonumber\\
		&3. V \text{ is a positive operator}    
	\end{align}
	Using equation \eqref{eqnho} we can show that the positivity condition in PT symmetric case becomes,
	\begin{equation}
		\braket{\bar R_{n}}{(O^\dagger)^{V}O|R_{n}}\geq 0
	\end{equation}
	where, $ \bra{\bar R_{m}}=\bra{ R_{m}}V$ and 
	$(O^\dagger)^{V}=V^{-1}O^{\dagger} V$. Using the following constraints we can obtain the eigenspectrum of any PT symmetric non-hermitian Hamiltonian, 
	\begin{align}\label{eqnhb}
		& \braket{\bar R_{n}}{[H,O] |R_{n}} =0\nonumber\\
		&\braket{\bar R_{n}}{HO |R_{n}} =E_{n}\braket{\bar R_{n}}{O |R_{n}}\nonumber\\
		&    \braket{\bar R_{n}}{(O^\dagger)^{V}O|R_{n}}\geq 0
	\end{align}
	
	In the next section we illustrate with few examples how to use these conditions to obtain Eigen spectrum of PT symmetric hamiltonians.
	\section{Example of PT symmetric potentials and implementation of bootstrap}\label{expt1}
	In this section we are going to find the eigenvalues of some PT symmetric Hamiltonian  using the above formalism .
	\subsection{($2\times 2 $) PT-symmetric Hamiltonian:}
	Let's consider the following ($2\times 2 $)
	PT symmetric non-hermitian Hamiltonian,
	\begin{equation}\label{eqsh}
		H=\begin{pmatrix}  i & \sqrt{2}  \\ \sqrt{2} & -i \end{pmatrix}
	\end{equation}
	For this Hamiltonian we find the $V$ operator and it is given by,
	\begin{equation}
		V=\begin{pmatrix}  \sqrt{2} & -i  \\ i & \sqrt{2} \end{pmatrix}
	\end{equation}
	Now to get the eigenvalues of $H$, let's consider the following operator,
	\begin{equation}
		O=a_{0}\mathbb{1} _{2\times 2}+a_{1}\sigma_{x}+a_{3}\sigma_{z}
	\end{equation}
	where, $a_{0}, a_{1}$ and $a_{3}$ are all real numbers. Using the equation \eqref{eqnhb} we get,
	\begin{equation}
		a^{2}_{0}+3 a^{2}_{1} +a^{2}_{2} +\sqrt{2}a_{0}a_{1}(E+1/E)+a_{1}a_{2}(E+1/E)\geq 0
	\end{equation}
	where, $E$ is the eigenvalue of $H$. The above inequality must hold for every values of  $a_{0}, a_{1}$ and $a_{3}$. We numirically find that the inequality holds for only $E=1.00$ and $E=-1.00$, so the possible eigenvalues of $H$ are $1.00$ and $-1.00$. The exact eigen values for this Hamiltonian are 1 and -1.
	\subsubsection{General ($2\times 2 $) PT-symmetric Hamiltonian:}
	Let's take the following general ($2\times 2 $)
	PT symmetric non-hermitian Hamiltonian,
	\begin{equation}
		H=\begin{pmatrix}  re^{i\theta} & s  \\ s &  re^{-i\theta} \end{pmatrix}
	\end{equation}
	where, $r, s$ and $\theta$ are all real number. The energy eigen values of this Hamiltonian are real as long as the following inequality holds, $s^2-r^2\sin^2{\theta}\geq 0$. The Hamiltonian defined in equation \eqref{eqsh} is a special case of the above Hamiltonian.
	For this Hamiltonian we find the $V$ operator and it is given by,
	\begin{equation}
		V=\frac{1}{\cos{\alpha}}\begin{pmatrix}  1 & -i\sin{\alpha}  \\ i\sin{\alpha} & 1 \end{pmatrix}
	\end{equation}
	where, $\sin{\alpha}=\frac{r}{s} \sin{\theta}$.
	Now to get the eigenvalues of $H$, let's consider the following operator,
	\begin{equation}
		O=a_{0}\mathbb{1} _{2\times 2}+a_{1}\sigma_{x}+a_{3}\sigma_{z}
	\end{equation}
	where, $a_{0}, a_{1}$ and $a_{3}$ are all real numbers. Using the equation \eqref{eqnhb} we get,
	\begin{equation}\label{eqdgp}
		a^{2}_{0}+3 a^{2}_{1}(1+\frac{2r^2\sin^2{\theta}}{s^2-r^2\sin^2{\theta}}) +a^{2}_{2} +a_{0}a_{1}e+a_{1}a_{2}\frac{r\sin{\theta}}{s}e\geq 0
	\end{equation}
	where, $e=s\frac{(s^2-r^2\sin^2{\theta})+(E-r\cos{\theta})^2}{(s^2-r^2\sin^2{\theta)}(E-r\cos{\theta})}$ and here  $E$ is the eigenvalue of $H$. The above inequality must hold for every values of  $a_{0}, a_{1}$ and $a_{3}$. We can now numirically find the eigenvalues of $H$ for arbitrary values of $r,s$ and $ \theta$. Here, we have given one example:
	Let's consider the following case where, $r=s=1 $ and $\theta=\frac{\pi}{4}$. We substitute these values in \eqref{eqdgp} and numirically find that the possible energy eigen values are 0.00 and 1.41. 
	The exact eigen values for this Hamiltonian are 0 and $\sqrt{2}$.

	\subsection{ Shifted Simple Harmonic Oscillator:}
	Let's consider the following PT symmetric non-hermitian Hamiltonian,
	\begin{equation}
		H=p^{2}+x^{2}+2i \epsilon x
	\end{equation}
	For this Hamiltonian the $V$ operator is \cite{swanson},
	\begin{equation}
		V=e^{2\epsilon p}
	\end{equation}
	Now under this operator $V$, $x$ and $p$
	changes in the following way,
	\begin{align}
		&x^{V}=V^{-1}x V=(x+2i\epsilon)\nonumber\\
		&p^{V}=V^{-1}pV=p
	\end{align}
	By choosing $O=\sum_{k}a_{k}p^{k}$ and using the positivity condition i.e. $\braket{\bar R_{n}}{(O^\dagger)^{V}O|R_{n}}\geq 0$, we got the following inequality,
	\begin{align}
		& \sum_{j,k}a^*_{j}a_{k}\braket{\bar R_{n}}{p^{j+k}|R_{n}}\geq 0
	\end{align}
	The above inequality implies the matrix, $\bar{M}_{jk}=\braket{\bar R_{n}}{p^{j+k}|R_{n}}$, should be positive semi-definite. Using equation \eqref{eqnhb} we find the following recursion relation,
	\begin{align}\label{eqr1}
		4t(E-\epsilon^2)\braket{\bar R_{n}}{p^{t-1}|R_{n}}+t(t-1)(t-2)\braket{\bar R_{n}}{p^{t-3}|R_{n}}-4(t+1)\braket{\bar R_{n}}{p^{t+1}|R_{n}}=0
	\end{align}
	Now using the positivity condition or $\bar{M}_{jk}$ is a positive semi definite matrix, we numirically find the eigenvalues of $H$. In this problem there is only one independent variable i.e. $E$, so the search space is $1D$. We have shown the ground and first excited state in two different plots because we have to take different matrix size of $\bar{M}_{jk}$ for the ground and excited state.
	\begin{figure}[h]
		\includegraphics[scale=0.6]{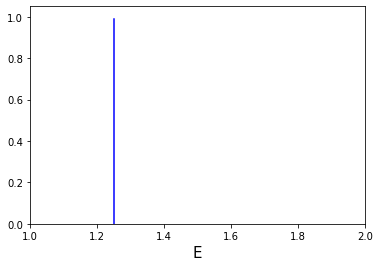} \includegraphics[scale=0.6]{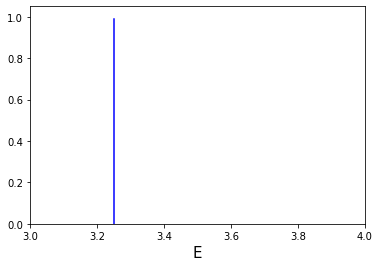}
		\caption{We have plotted the ground state and the first excited state energy eigen values of the shifted harmonic oscillator at $\epsilon=0.5$. The exact ground state and the first excited state energy values of the shifted harmonic oscillator at $\epsilon=0.5$ are, $E=1.25$ and $E=3.25$ respectively.}
		\label{fig:my_label}
	\end{figure}
	\begin{figure}[!]
		\includegraphics[scale=0.6]{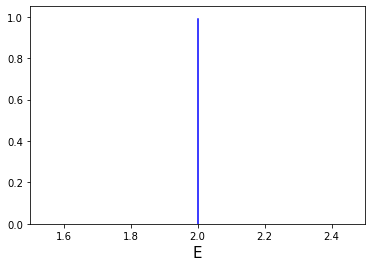} \includegraphics[scale=0.6]{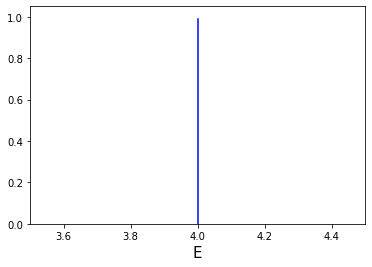}
		\caption{We have plotted the ground state and the first excited state energy eigen values of the shifted harmonic oscillator at $\epsilon=1$. The exact ground state and the first excited state energy values of the shifted harmonic oscillator at $\epsilon=1$ are, $E=2$ and $E=4$ respectively.}
		\label{fig:my_label}
	\end{figure}

	\newpage
	
	\subsection{ Swanson Hamiltonian:}
	Here we consider the non-Hermitian  Swanson Hamiltonian \cite{swanson} which has the following form,
	
	\begin{equation}
		H=p^{2}+x^{2}+ic(xp+px)
	\end{equation}
	The $V$ operator for this Hamiltonian is  \cite{swanson},
	\begin{equation}
		V=e^{-cx^2}
	\end{equation}
	Now under this operator $V$, $x$ and $p$
	changes in the following way,
	\begin{align}
		&x^{V}=V^{-1}x V=x\nonumber\\
		&p^{V}=V^{-1}pV=(p+2icx)
	\end{align}
	By choosing $O=\sum_{k}b_{k}x^{k}$ and using the positivity condition i.e. $\braket{\bar R_{n}}{(O^\dagger)^{V}O|R_{n}}\geq 0$, we got the following inequality,
	\begin{align}
		& \sum_{j,k}b^*_{j}b_{k}\braket{\bar R_{n}}{x^{j+k}|R_{n}}\geq 0
	\end{align}
	The above inequality implies the matrix, $\bar{M}_{jk}=\braket{\bar R_{n}}{x^{j+k}|R_{n}}$, should be positive semi-definite. Using equation \eqref{eqnhb} we find the following recursion relation,
	\begin{align}\label{eqr2}
		4tE\braket{\bar R_{n}}{x^{t-1}|R_{n}}+t(t-1)(t-2)\braket{\bar R_{n}}{x^{t-3}|R_{n}}-4(1+c^2)(t+1)\braket{\bar R_{n}}{x^{t+1}|R_{n}}=0
	\end{align}
	Now using the positivity condition or $\bar{M}_{jk}$ is a positive semi definite matrix, we numirically find the eigenvalues of $H$. For this Hamiltonian also the search space is 1D.
	\begin{figure}[h]
		\includegraphics[scale=0.5]{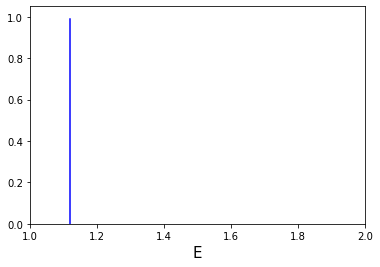} \includegraphics[scale=0.5]{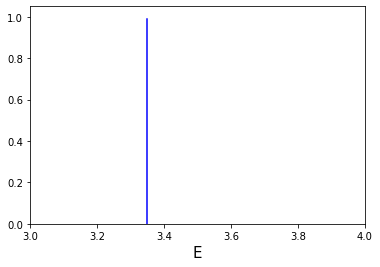}
		\caption{We have plotted the ground state and the first excited state energy eigen values of the Swanson Hamiltonian oscillator at $\epsilon=0.5$. The exact ground state and the first excited state energy values of the Swanson Hamiltonian at $\epsilon=0.5$ are, $E=1.12$ and $E=3.35$ respectively.}
		\label{fig:my_label2}
	\end{figure}
	\begin{figure}[h]
		\includegraphics[scale=0.5]{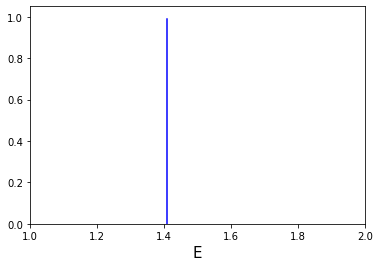} \includegraphics[scale=0.5]{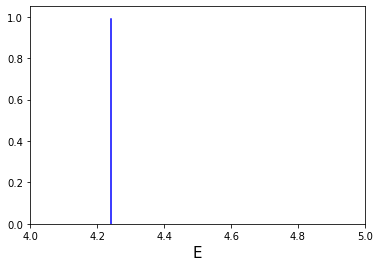}
		\caption{We have plotted the ground state and the first excited state energy eigen values of the Swanson Hamiltonian at $\epsilon=1$. The exact ground state and the first excited state energy values of the Swanson Hamiltonian at $\epsilon=1$ are, $E=1.41$ and $E=4.24$ respectively.}
		\label{fig:my_label1}
	\end{figure}

	\newpage
	
	\subsection{ PT-symmetric Pöschl–Teller potential:}
	The Hamiltonian for PT-symmetric Pöschl–Teller potential is given by,
	\begin{equation}\label{eqptp}
		H=\frac{p^2}{2}- \frac{\lambda(\lambda+1)}{2} \sech^2{\big(x+i\epsilon\big)}
	\end{equation}
	For this Hamiltonian the $V$ operator is,
	\begin{equation}
		V=e^{2\epsilon p}
	\end{equation}
	Now under this operator $V$, $x$ and $p$
	changes in the following way,
	\begin{align}
		&x^{V}=V^{-1}x V=(x+2i\epsilon ) \nonumber\\
		&p^{V}=V^{-1}pV=p
	\end{align}
	Now let's define, $z=x+i\epsilon$ and substitute this in equation \eqref{eqptp} then the equation \eqref{eqptp} becomes,
	\begin{equation}
		H=\frac{p^2}{2}- \frac{\lambda(\lambda+1)}{2} \sech^2\big({z}\big)
	\end{equation}
	We can easily show that, $z$ has the following property, 
	\begin{align}
		(z^\dagger)^{V}=V^{-1}(x-i\epsilon) V=z
	\end{align}
	By choosing $O=\sum_{k}b_{k}\sech^{2k}{\big(z\big)}$ and using the positivity condition i.e. $\braket{\bar R_{n}}{(O^\dagger)^{V}O|R_{n}}\geq 0$, we got the following inequality,
	\begin{align}
		\sum_{j,k}b^*_{j}b_{k}\braket{\bar R_{n}}{\sech^{2[j+k]}{\big( z \big)}\big|R_{n}}\geq 0
	\end{align}
	The above inequality implies the matrix, $\bar{M}_{jk}=\braket{\bar R_{n}}{\sech^{2[j+k]}\big|R_{n}}$, should be positive semi-definite. Using equation \eqref{eqnhb} we find the following recursion relation,
	\begin{align}
		&(2tE+\frac{t^3}{2})\braket{\bar R_{n}}{\sech^{t}{\big(z\big)}|R_{n}}+\big(\frac{t^3}{2}+3t^2+4t+2+\frac{3t+2}{2}-2\lambda(\lambda+1)(t+2)\big)\braket{\bar R_{n}}{\sech^{t+2}{\big(z\big)}|R_{n}}\nonumber\\
		&+\big(-2(t+1)E+2\lambda(\lambda+1)(t+1)-t^3-3t^2-4t-2\big)\braket{\bar R_{n}}{\sech^{t+4}{\big(z\big)}|R_{n}}=0
	\end{align}

	Now using the positivity condition or $\bar{M}_{jk}$ is a positive semi definite matrix, we numirically find the eigenvalues of $H$. This potential is exactly solvable whose eigenvalues  The plots obtained after bootstrap are shown in Fig \ref{fig:plot1}

	\subsection{ $p^2-x^4$ potential:}
	Let's consider the following Hamiltonian,
	\begin{equation}\label{eqp5.0}
		H=p^2-x^4
	\end{equation}
	The potential $V=-x^4$ is unbounded below on the real line but if we take $x$ on a contour in the lower-half complex plane  then it can give rise to a well-posed
	bound state problem. However, the potential becomes 
	PT-symmetric rather than Hermitian \cite{-x4,benderprl} and the equivalent PT-symmetric Hamiltonian is given by, 
	\begin{equation}\label{eqp5}
		H=\frac{1}{2}\{(1+ix),p^{2}\}-\frac{1}{2}p-\alpha (1+ix)^2
	\end{equation}
	where, $\alpha=16$. The $V$ operator for this Hamiltonian is \cite{-x4},
	\begin{equation}
		V=e^{(\frac{p^3}{3\alpha}-2p)}
	\end{equation}
	Now under this operator $V$, $x$ and $p$
	changes in the following way,
	\begin{align}
		&x^{V}=V^{-1}x V=x+i(\frac{p^2}{\alpha}-2)\nonumber\\
		&p^{V}=V^{-1}pV=p
	\end{align}
	By choosing $O=\sum_{k}a_{k}p^{k}$ and using the positivity condition i.e. $\braket{\bar R_{n}}{(O^\dagger)^{V}O|R_{n}}\geq 0$, we got the following inequality,
	\begin{align}
		& \sum_{j,k}a^*_{j}a_{k}\braket{\bar R_{n}}{p^{j+k}|R_{n}}\geq 0
	\end{align}
	The above inequality implies the matrix, $\bar{M}_{jk}=\braket{\bar R_{n}}{p^{j+k}|R_{n}}$, should be positive semi-definite. Using equation \eqref{eqnhb} we find the following recursion relation,
	\begin{align}
		&4\alpha t E\braket{\bar R_{n}}{p^{t-1}|R_{n}}+(2t+1)\alpha \braket{\bar R_{n}}{p^{t}|R_{n}}+\alpha^2 t(t-1)(t-2)\braket{\bar R_{n}}{p^{t-3}|R_{n}}\nonumber\\
		&-(t+2)\braket{\bar R_{n}}{p^{t+3}|R_{n}}=0
	\end{align}
	Now using the positivity condition or $\bar{M}_{jk}$ is a positive semi definite matrix, we numirically find the eigenvalues of $H$. A similar analysis can be done for Hamiltonian of the form $H=p^2+m^2 x^2-x^4$ which can be mapped to $\frac{1}{2}\{(1+ix),p^{2}\}-\frac{1}{2}p-m^2(1+ix)-\alpha (1+ix)^2$. 
	\begin{figure}[h]
		\centering
		\includegraphics[scale=0.75]{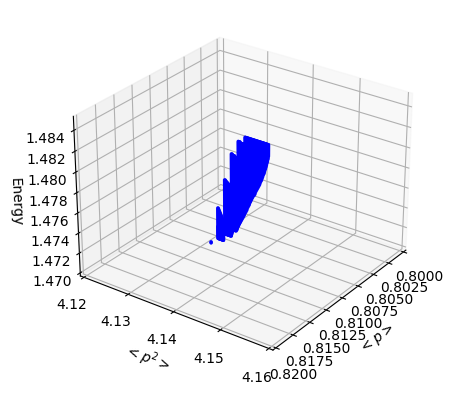} 
		\caption{We have plotted the ground state and the first excited state energy eigen values of eq \ref{eqp5.0}}
		\label{fig:plot3}
	\end{figure}
	\begin{figure}[h!]
	 	\centering
   	    \includegraphics[scale=0.75]{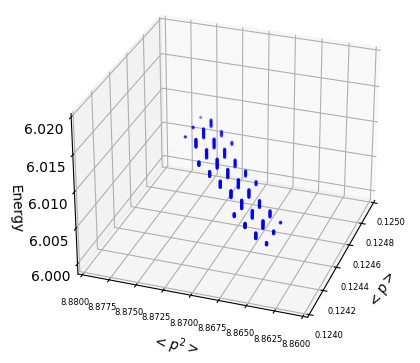}
		\caption{We have plotted the ground state and the first excited state energy eigen values of eq \ref{eqp5.0}}
		\label{fig:plot3}
	\end{figure}

\subsection{Simple Harmonic Oscillator Coupled to a Shifted Simple Harmonic Oscillator}\label{s7}
	The Hamiltonian of the two coupled PT-symmetric harmonic oscillator is given by,
	\begin{equation}\label{eqp7}
		H=p^{2}+x^{2}+q^{2}+y^{2}+2i y+2\epsilon xy
	\end{equation}
	The $V$ operator for this Hamiltonian is 
	\begin{equation}
		V=e^{-2(\alpha p+\beta q)}
	\end{equation}
	where, 
	\begin{align}
		\alpha =\frac{\epsilon}{1-\epsilon^2}, \;\;\;\;\;\;\;\;\;\beta =-\frac{1}{1-\epsilon^2}
	\end{align}
	Now under this operator $V$,  $p$ and $q$ does not change but $x$ and $y$
	changes in the following way,
	\begin{align}
		&x^{V}=V^{-1}x V=(x-2i\alpha)\nonumber\\
		&y^{V}=V^{-1}yV=(y-2i\beta)
	\end{align}
	Now let's define, $x_{1}=x-i\alpha$, $y_{1}=y-i\beta$ and substitute this in equation \eqref{eqp7} then the equation \eqref{eqp7} becomes,
	\begin{equation}
		H=p^{2}+x_{1}^{2}+q^{2}+y_{1}^{2}+2\epsilon x_{1}y_{1}+\frac{1}{1-\epsilon^2}
	\end{equation}
	We can easily show that, $x_{1}$ and $y_{1}$ has the following property, 
	\begin{align}
		(x_{1}^\dagger)^{V}=V^{-1}(x+i\alpha) V=x_{1}\nonumber\\
		(y_{1}^\dagger)^{V}=V^{-1}(y+i\beta) V=y_{1}
	\end{align}

\begin{figure}[h!]
		\centering
		\includegraphics[scale=0.6]{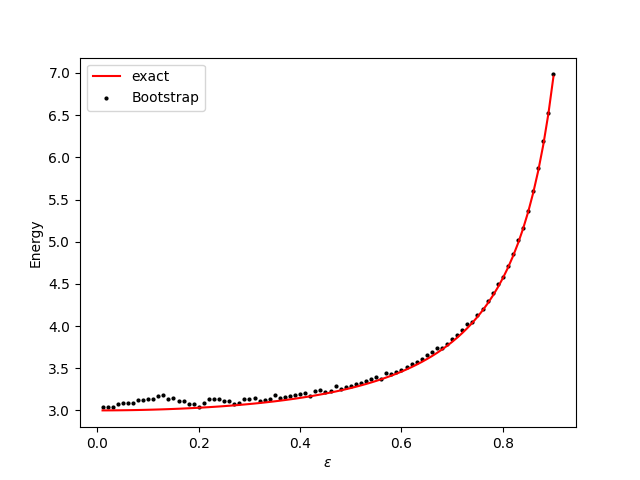} 
		\caption{Here we plot the results of bootstrap for $H=p^{2}+x_{1}^{2}+q^{2}+y_{1}^{2}+2\epsilon x_{1}y_{1}+\frac{1}{1-\epsilon^{2}}$ and compare it against the exact spectrum. The exact energy eigenvalue solution which is plotted in red line in the above figure is given by E (ground state)=$\sqrt{1+\epsilon}+\sqrt{1-\epsilon}+\frac{1}{1-\epsilon^{2}}$}
		\label{fig:plot2}
	\end{figure}

	For this two-dimensional problem it is difficult to find a closed form recursion relation like \eqref{eqr1} or \eqref{eqr2}. However, in the Ref. \cite{hartnol},  Xizhi Han,  Sean A. Hartnoll, etc. explained how to deal with this type of problem. First consider the following trial operators $I, x_{1},y_{1},p$, and $q$. From the positivity condition defined in equation \eqref{eqnhb}, the following bootstrap matrix should be positive semi definite and this condition is used to get Fig\ref{fig:plot2}:
	
	\begin{table}[h!]
		\begin{center}
			
			\begin{tabular}{c|c c c c c}
				\textbf{}&$I$&$x_{1}$&$y_{1}$&$p$&$q$\\
				\hline
				$ I$ & $1$ &$0$&$0$&$0$&$0$\\
				$ x_{1}$ & $0$ &$\langle x^{2}_{1} \rangle$&$\frac{\langle p^{2} \rangle - \langle x^{2}_{1} \rangle}{\epsilon}$&$\frac{i}{2}$&$0$\\
				$ y_{1}$ & $0$ &$\frac{\langle p^{2} \rangle - \langle x^{2}_{1} \rangle}{\epsilon}$&$\langle x^{2}_{1} \rangle$&$0$&$\frac{i}{2}$\\
				$ p$ & $0$ &$\frac{-i}{2}$&$0$&$\langle p^{2} \rangle$&$ \frac{\langle p^{2} \rangle - (\epsilon^{2} -1) \langle x^{2}_{1} \rangle}{\epsilon}$\\
				$ q$ & $0$ &$0$&$\frac{-i}{2}$&$ \frac{\langle p^{2} - (\epsilon^{2} -1) \langle x^{2}_{1} \rangle}{\epsilon}$&$\langle p^{2} \rangle$\\		
			\end{tabular}
			\caption{Matrix for Bootstrapping  $H=p^{2}+x_{1}^{2}+q^{2}+y_{1}^{2}+2\epsilon x_{1}y_{1}+\frac{1}{1-\epsilon^{2} } $}
		\end{center}
		\label{tab:t2}
	\end{table}

	\vspace{1mm}

	\subsection{Simple Harmonic Oscillator Coupled to Swanson Hamiltonian}\label{s8}
	Let's consider the following PT-symmetric non hermitian two-dimensional potential,
	\begin{equation}\label{eqp8}
		H=p^{2}+x^{2}+q^{2}+y^{2}+ic\{q,y\}+2\epsilon xy
	\end{equation}
	The $V$ operator for this Hamiltonian is 
	\begin{equation}
		V=e^{-cy^2}
	\end{equation}
	Now under this $V$ operator,  $x$, $y$ and $p$ does not change but $q$ 
	changes in the following way,
	\begin{align}
		&q^{V}=V^{-1}q V=(q+2icy)
	\end{align}
	Now let's define, $q_{1}=q+icy$ and substitute this in equation \eqref{eqp8} then the equation \eqref{eqp8} becomes,
	\begin{equation}
		H=p^{2}+x^{2}+q_{1}^{2}+(1+c^2)y^{2}+2\epsilon xy
	\end{equation}
	We can easily show that, $q_{1}$ has the following property, 
	\begin{align}
		(q_{1}^\dagger)^{V}=V^{-1}(q-icy) V=q_{1}
	\end{align}
	For this two-dimensional problem it is difficult to find a closed form recursion relation like \eqref{eqr1} or \eqref{eqr2}. However, in the Ref. \cite{hartnol},  Xizhi Han,  Sean A. Hartnoll, etc. explained how to deal with this type of problem. First consider the following trial operators $I, x,y,p$, and $q_{1}$. From the positivity condition defined in equation \eqref{eqnhb}, the following bootstrap matrix should be positive semidefinite and this condition is used to get Fig\ref{fig:plot3}:
	
	\begin{table}[h!]
		\tiny
		\begin{center}
			
			\begin{tabular}{c|c c c c c}
				\textbf{}&$I$&$x$&$y$&$p$&$q_{1}$\\
				\hline
				$ I$ & $1$ &$0$&$0$&$0$&$0$\\
				$ x$ & $0$ &$\langle x^{2} \rangle$&$\frac{\langle p^{2} \rangle - \langle x^{2} \rangle}{\epsilon}$&$\frac{i}{2}$&$0$\\
				$ y$ & $0$ &$\frac{\langle p^{2} \rangle - \langle x^{2} \rangle}{\epsilon}$&$\langle p^{2} \rangle \left(\frac{\alpha-1}{\epsilon^{2}} \right) + \langle x^{2} \rangle \left(1-\frac{\alpha -1}{\epsilon^{2}} \right) $&$0$&$\frac{i}{2}$\\
				$ p$ & $0$ &$\frac{-i}{2}$&$0$&$\langle p^{2} \rangle$&$ \langle p^{2} \rangle \frac{\alpha}{\epsilon}+\langle x^{2} \rangle \left(\epsilon - \frac{\alpha}{\epsilon} \right)$\\
				$ q_{1}$ & $0$ &$0$&$\frac{-i}{2}$&$ \langle p^{2} \rangle \frac{\alpha}{\epsilon}+\langle x^{2} \rangle \left(\epsilon - \frac{\alpha}{\epsilon} \right)$&$\langle p^{2} 
				\rangle\left(1-\frac{\alpha(\alpha-1)}{\epsilon^{2}}\right) +
				\langle x^{2} \rangle \left(\alpha -\left(1-\frac{\alpha(\alpha-1)}{\epsilon^{2}}\right) \right)   $\\		
			\end{tabular}
			\caption{Matrix for Bootstrapping $H=p^{2}+x^{2}+q_{1}^{2}+(1+c^2)y^{2}+2\epsilon xy$}
		\end{center}
	\end{table}

	\begin{figure}[h!]
		\centering
		\includegraphics[scale=0.70]{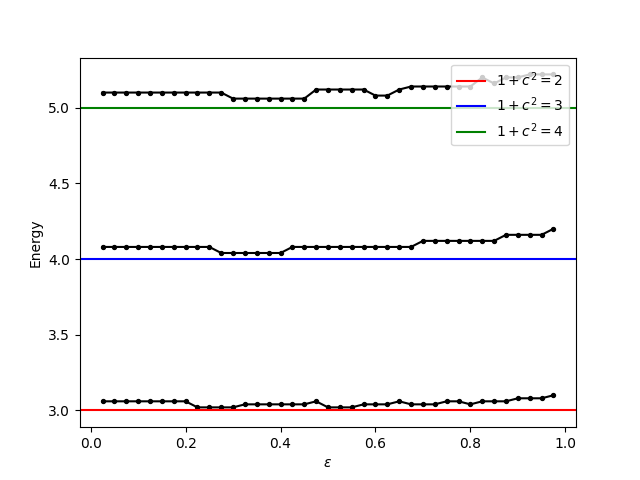} 
		\caption{Here we plot the results of bootstrap for $H=p^{2}+x^{2}+q_{1}^{2}+(1+c^2)y^{2}+2\epsilon xy$ and compare it against the exact spectrum. The exact ground state energy for this system is given by $1+\alpha$ where $\alpha=1+c^{2}$; hence it is a constant with respect to $\epsilon$. We have plotted for 3 different values of $\alpha$. The scatter plot in black is the actual bootstrap plot and the lines in red blue and green are the constant energy eigenvalues}
		\label{fig:plot3}
	\end{figure}
	\newpage
	\section{Discussion} \label{disc1}
	Bootrapping in quantum mechanics is the use of positivity condition to obtain spectrum. However for non-hermitian systems, the usual positivity condition does not work. In this paper we deal with particular class of non-hermitian hamiltonian, the PT symmetric hamiltonian for which case the eigen spectrum is real.  In this paper we point out how to  generalise bootstrap  technique for PT symmetric case. We show that   one can define a modified positivity condition and use it to solve for the Eigen spectrum.  We illustrate this with several examples.
	
	One of the important application of the PT symmetric Hamiltonian is to describe the gain-loss system. The gain-loss system consists of two subsystem and it is not an isolated system because it is in contact with external environment. When this system is in dynamical equilibrium i.e. loss and gain are equal, it exhibits PT symmetry i.e. the Hamiltonian of the composite system is PT symmetric. 
	There are also many situation where PT symmetry\footnote{The ground state of a Bose system of hard spheres is described by a non-Hermitian Hamiltonian \cite{PhysRev.115.1390}} comes out naturally like in studies of
	the Lee-Yang edge singularity \cite{PhysRevLett.40.1610,CARDY1989275}. Our study of PT symmetric potentials motivates us to develop bootstrapping for more general class of non-hermitian hamiltonians. This might have potential application on quantum open systems \cite{breuer}. One important goal would be to see if one can obtain the spectrum of the Lindbladian operator.  This development will have far reaching applications such as in transport phenomenon and bootstrapping may be used to derive various bounds. We shall report on these exciting possibilities in future.

	\textbf{Acknowledgements} 
	Work of SJ is supported by Ramanujan Fellowship. Work of SK is supported by CSIR fellowship with Grant Number 09/0936(11643)/2021-EMR-I. Y. Agarwal is supported by KVPY fellowship. SJ would like to thank S.J.Ganesh for helpful discussions. SK would like to thank L. Bhandari and S. Pande for discussions. YA,DT would like to thank A. Ravishankar for discussions.The authors would also like to thank the people of India for their steady support in basic research. 
	
	\appendix
	\section{Some details}\label{detail}
	In this appendix, we describe the method used to bootstrap potentials in \ref{s7} and \ref{s8} First, a matrix is formed as mentioned in Table \ref{tab:t1}. Then $\langle[\mathcal{H},\mathcal{O}]\rangle=0$ identity is used to establish constraints on the expectation values of observables,  \newline ($\mathcal{O}=\{I,X,Y,P,Q,XP,PX,......\}$). These relations are used to find out the independent variables and rewrite the matrix in Table\ref{tab:t1} in terms of the independent variables. For example, for the potential in \ref{s7}, all other expectation values were written in terms of $\langle x^{2}\rangle$ and $\langle p^{2}\rangle$; this is shown in Table\ref{tab:t2}. Next, a simple program  searches for different values of  $\langle x^{2}\rangle$ and $\langle p^{2}\rangle$ for which the matrix in Table 2 is positive semi-definite and minimises the expression for ground state energy of the Hamiltonian. A similar computation was done for Sec \ref{s8} as well.
		\begin{table}[h!]
		\begin{center}
			
			\begin{tabular}{l|c r r r r}
				\textbf{}&$I$&$x$&$y$&$p$&$q$\\
				\hline
				$ I$ & $1$ &$\langle x \rangle$&$\langle y \rangle$&$\langle p \rangle$&$\langle q \rangle$\\
				$ x$ & $\langle x \rangle$ &$\langle x^{2} \rangle$&$\langle xy \rangle$&$\langle xp \rangle$&$\langle xq \rangle$\\
				$ y$ & $\langle y \rangle$ &$\langle yx \rangle$&$\langle y^{2} \rangle$&$\langle yp \rangle$&$\langle yq \rangle$\\
				$ p$ & $\langle p \rangle$ &$\langle px \rangle$&$\langle py \rangle$&$\langle p^{2} \rangle$&$\langle pq \rangle$\\
				$ q$ & $\langle q \rangle$ &$\langle qx \rangle$&$\langle qy \rangle$&$\langle qp \rangle$&$\langle p^{2} \rangle$\\		
			\end{tabular}
			\caption{General Matrix for Bootstrapping. This matrix will be used in Sec \ref{s7} and Sec \ref{s8} where a good recursion relation cannot be obtained}
		\end{center}
		\label{tab:t1}
	\end{table}
	
	\newpage
	
	\bibliography{abc.bib}

\bibliographystyle{JHEP}
\end{document}